\documentclass[sigconf]{acmart}
\usepackage{todonotes}
\usepackage[acronym]{glossaries}
\usepackage[capitalize]{cleveref}
\usepackage[inline]{enumitem}
\usepackage{layout}
\usepackage{url}

\newcommand{\appname}{BTLEmap}
\AtBeginDocument{%
  \providecommand\BibTeX{{%
    \normalfont B\kern-0.5em{\scshape i\kern-0.25em b}\kern-0.8em\TeX}}}

\copyrightyear{2020} 
\acmYear{2020} 
\setcopyright{rightsretained} 

\acmConference[WiSec '20]{WiSec '20: ACM Conference on Security and Privacy in Wireless and Mobile Networks}{June 09--10, 2020}{Linz (Virtual Event), Austria}
\acmBooktitle{WiSec '20: ACM Conference on Security and Privacy in Wireless and Mobile Networks, June 08--10, 2020, Linz (Virtual Event), Austria}

\begin{document}

\title{DEMO: \appname: Nmap for Bluetooth Low Energy}


\author{Alexander Heinrich}
    \affiliation[obeypunctuation=true]{
        \department[0]{Secure Mobile Networking Lab}\\
        \department[1]{Department of Computer Science}\\
        \institution{TU Darmstadt}, \country{Germany}
    }
    \email{aheinrich@seemoo.de}
    \orcid{0000-0002-1150-1922}
    
\author{Milan Stute}
    \affiliation[obeypunctuation=true]{
            \department[0]{Secure Mobile Networking Lab}\\
            \department[1]{Department of Computer Science}\\
            \institution{TU Darmstadt}, \country{Germany}
        }
    \email{mstute@seemoo.de}
    \orcid{0000-0003-4921-8476}

\author{Matthias Hollick}
    
    \affiliation[obeypunctuation=true]{
            \department[0]{Secure Mobile Networking Lab}\\
            \department[1]{Department of Computer Science}\\
            \institution{TU Darmstadt}, \country{Germany}
        }
    \email{mhollick@seemoo.de}
    \orcid{0000-0002-9163-5989}


\renewcommand{\shortauthors}{Heinrich and Stute}

\begin{abstract}
The market for \gls{BLE} devices is booming and, at the same time, has become an attractive target for adversaries.
To improve \gls{BLE} security at large, we present \appname{}, an auditing application for \gls{BLE} environments.
\appname{} is inspired by network discovery and security auditing tools such as Nmap for IP-based networks. It allows for device enumeration, \gls{GATT} service discovery, and device fingerprinting. It also features a \gls{BLE} advertisement dissector, data exporter, and a user-friendly UI including a proximity view. \appname{} currently runs on iOS and macOS using Apple's CoreBluetooth API but also accepts alternative data inputs such as a Raspberry Pi to overcome the restricted vendor API.
The open-source project is under active development and will provide more advanced capabilities such as long-term device tracking (in spite of MAC address randomization) in the future.

\end{abstract}


\acmConference[WiSec '20]{13th ACM Conference on Security and Privacy in Wireless and Mobile Networks}{July 8--10, 2020}{Linz (Virtual Event), Austria}
\acmBooktitle{13th ACM Conference on Security and Privacy in Wireless and Mobile Networks (WiSec '20), July 8--10, 2020, Linz (Virtual Event), Austria}\acmDOI{10.1145/3395351.3401796}
\acmISBN{978-1-4503-8006-5/20/07}



\maketitle


\glsdisablehyper 

\newacronym{BLE}{BLE}{Bluetooth Low Energy}
\newacronym{RSSI}{RSSI}{Received signal strength indication}
\newacronym{GATT}{GATT}{Generic Attribute Profile}
\newacronym{OS}{OS}{operating system}
\newacronym{AoA}{AoA}{angle-of-arrival}
\newacronym{AWDL}{AWDL}{Apple Wireless Direct Link}

\glsresetall

\begin{figure}
    \centering
    \includegraphics[width=\linewidth]{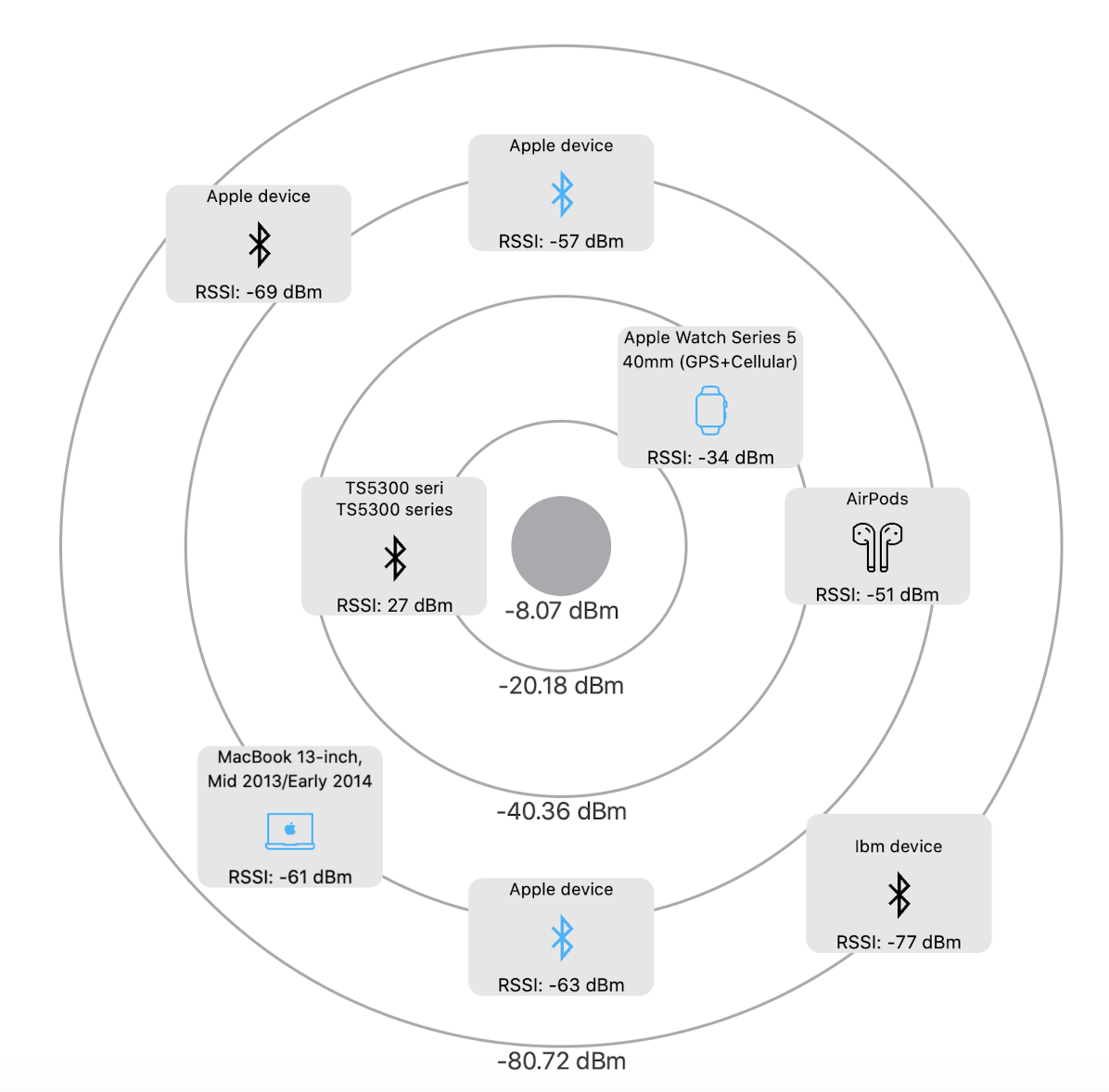}
    \caption{\appname{}'s proximity view}
    \label{fig:environment_scanner}
\end{figure}

\section {Introduction and Background}


Since the introduction of \gls{BLE} in 2010~\cite{blCore4_0}, the technology has become widely adopted in smartphones, wearables, and other IoT devices.
In a modern household, there can be more than a dozen devices that supporting this technology. They are constantly sending \gls{BLE} advertisements to inform devices in their surrounding about their presence but typically without the owner noticing.
Most recently, \gls{BLE} has been proposed to be used for contact tracing during the COVID-19 pandemic. Most of the tracing solutions use \gls{BLE} advertisements containing a custom identification token that can be linked to an infected person.

\gls{BLE} advertisements pose a vast surface for privacy-com\-pro\-mising attacks. In the past, several researchers found that \gls{BLE} advertisements may contain fixed identifiers that allow for device tracking despite MAC address randomization~\cite{martin2019handoff} and could contain personally identifiable information about the user such as their phone number or email address~\cite{stute_airdrop}.
Besides, an adversary might place a tracking device based on \gls{BLE}, such as the rumored Apple \textit{AirTags}~\cite{AirTags}, in a person's pocket and leverage a crowd-sourced finder network to track their target.
In essence, \gls{BLE} devices share potentially sensitive data with devices in proximity. To improve our understanding of the (privacy-related) attack surface at large, e.\,g., by analyzing privacy leaks or detecting malicious devices, we as a security community require better application support.

We propose \appname{} (pronounce as \emph{beetle map}), a network discovery and security auditing tool in the spirit of Nmap~\cite{nmap} but for \gls{BLE} environments.
Currently, \appname{} supports device enumeration, advertisement dissection, rudimentary device fingerprinting, a proximity view, and more.
With its extensible and modular design, it will support more advanced features such as long-term device tracking and comprehensive fingerprinting capabilities in the future.
We make \appname{} for iOS and macOS available as open-source software on GitHub and distribute the application bundle for iOS and macOS via Apple's official App Store for free.

%
%

\section{\appname{}}

\begin{figure}
    \centering
    \includegraphics[width=\linewidth]{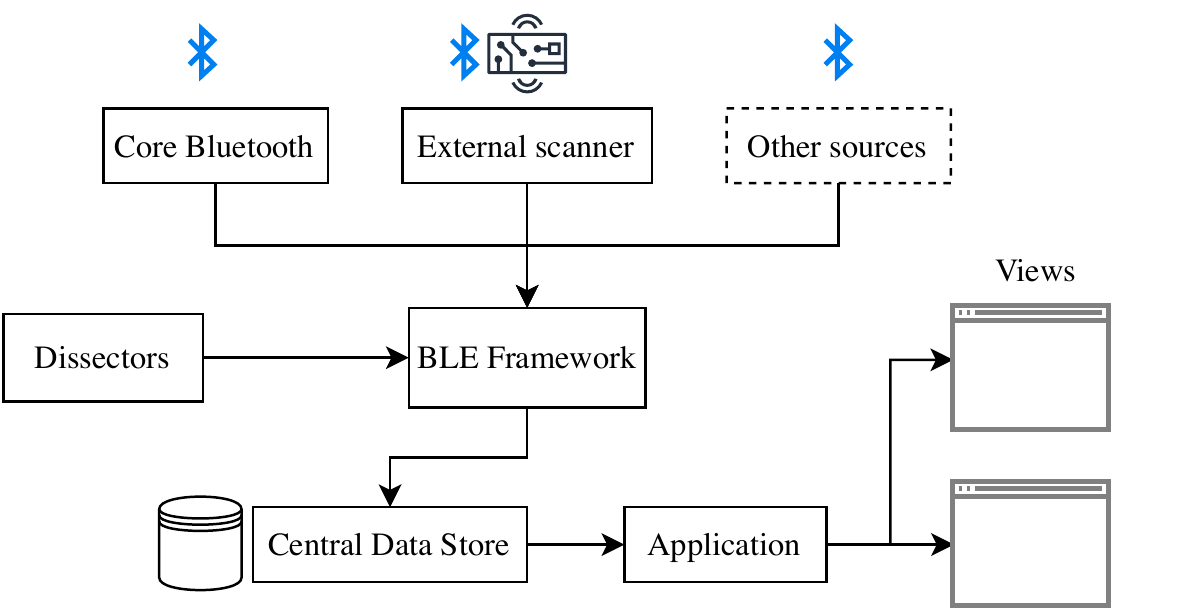}
    \caption{\appname's software architecture}
    \label{fig:architecture}
\end{figure}

Our initial version of \appname{} runs on iOS and macOS using SwiftUI and macOS Catalyst to enable rich animations combined with a unified user experience across multiple platforms. 
Our tool should be the first approach to analyze surroundings for \gls{BLE} devices and for researchers who want to discover new targets that should be analyzed. It automatically scans the environment for \gls{BLE} devices, lists their advertisements, connects to them to detect the supported services. All received advertisements are stored for long term analysis and can be exported. 

Our design favors modularity, as shown in \cref{fig:architecture}. At its core, \appname{} uses a data store for recording advertisements. It supports both internal and external input sources (more in \cref{sec:raspberry_pi}) and an extensible dissector module that already implements multiple dissectors for Apple's \gls{BLE} advertisements.
Finally, different views, such as a proximity view (\cref{sec:env_scanner}), present the advertisement data in different ways.
In the following, we highlight some of \appname{}'s features, in particular, the proximity view, the advertisement dissector, the automated device detection, and support for external scanning sources. Also, we discuss areas for future work.

\subsection{Proximity view}
\label{sec:env_scanner}

Using the central data store, \appname{} can create different representations of the \gls{BLE} readings. 
\Cref{fig:environment_scanner} shows a proximity view that estimates the distance to devices by using the \gls{RSSI} values in combination with the advertised transmission power of each advertisement that has been received.
The scanner can be used to discover devices in proximity quickly and to estimate how far away devices are. Even though \gls{RSSI} values cannot be used to measure an exact distance, it is possible to estimate a relative distance between the devices shown on the circular plane. 
At the moment, the angles shown are random because the current chips and APIs do not allow to measure the \gls{AoA}. As this feature is part of the current Bluetooth Core Specification \cite[1.A.8]{blCore5_2}, we plan to implement it as soon as \gls{AoA} becomes available to the public.

\subsection{Advertisement dissector}

\gls{BLE} advertisements contain up to 31 bytes of information about the emitting device or its services.
Existing \gls{BLE} scanners such as Nordic's nRF Connect~\cite{nRFConnect} display this information as a simple byte string as there is no standard on what information can be encoded or how it is encoded.
\appname{} leverages the reverse engineering efforts of several researcher teams~\cite{martin2019handoff,Celosia2020,stute_airdrop} to present the user with additional information about the state of the device. For example, Apple AirPods indicate the model name, colors, and battery status. We depict our Wireshark-inspired dissector view in \cref{fig:airpods_ble}.

\subsection{Automated device detection}
Apart from analyzing \gls{BLE} advertising packets, \appname{} has the option to automatically connect to all devices in range, query the supported services, and read the values of characteristics. 
This results in automated device detection, listing the manufacturer, device type, device names, and more. Many users tend to rename their Bluetooth enabled speakers or headphones, which can result in a privacy leak and tracking possibilities because \gls{BLE} often allows unauthenticated connections. \appname{} can be used to make such issues visible for researchers and for non-scientific users. 

\subsection{External scanning sources}
\label{sec:raspberry_pi}

The iOS and macOS APIs for \gls{BLE} scanning are limited. We have found that the \textit{CoreBluetooth} API on iOS does not report any manufacturer data that starts using Apple's company identifier \texttt{0x4c00} (little-endian). On macOS, the manufacturer data is not stripped, but the API will not retrieve all \gls{GATT} services supported by a device. The frameworks strip away Apple-specific services that should not be accessible from other devices.
Furthermore, it is not possible to retrieve a device's (randomized) \gls{BLE} MAC address, because \textit{bluetoothd} replaces all MAC addresses with an on-demand generated UUID.

\begin{figure}
    \centering
    \includegraphics[width=\linewidth]{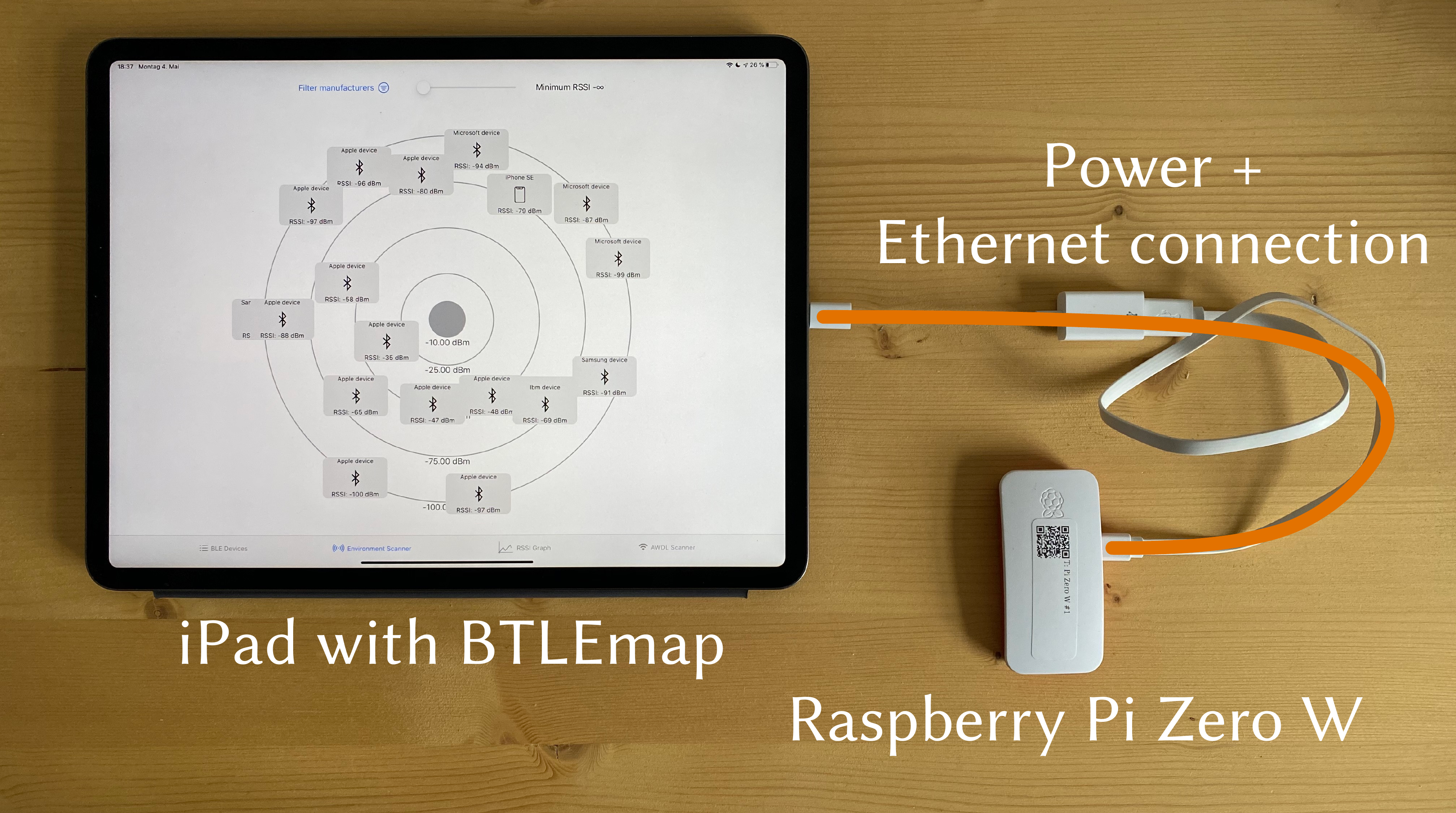}
    \caption{\appname{} setup with an external scanner source}
    \label{fig:external_scanner}
\end{figure}

These inherent restrictions would limit the usefulness of \appname{}. Therefore, we support external Bluetooth scanner inputs such as a Raspberry Pi (3 or newer/Zero W). The external device implements a simple Bluetooth scanner and forwards all received advertisements and discovered services in real-time to a connected macOS or iOS device. The Raspberry Pi connects over a direct Ethernet connection or by using Ethernet-over-USB to receive power and set up a connection, as shown in \cref{fig:external_scanner}. \appname{} announces itself as services via multicast DNS such that the Raspberry Pi can automatically connect to it.
The modular setup allows to support most functionality on the device running the app, but increase the feature spectrum by adding additional hardware.

\subsection{Additional features}

\appname{} currently supports several additional features: 
\begin{enumerate*}
    \item An \gls{RSSI} graph displaying all devices and the received RSSI values at a specific time. 
    \item An \gls{RSSI} recorder that allows recording RSSI readings and exporting them to CSV files later.
    \item Several filters, such as manufacturer or RSSI.
    \item Highlighting the recently active devices.
    \item Import and export of \texttt{pcap} files.
\end{enumerate*}

\subsection{Future work}
\label{sec:future_work}

At the moment, we are fine-tuning our implementation to be ready for a public release. Additionally, we work on algorithms that allow further automatic analysis of \gls{BLE} advertisements by identifying ``trackable'' identifiers, thus, revealing potentially private data. 

We started with macOS and iOS, because it allows us to combine \gls{BLE} with Apple's custom protocol \gls{AWDL} \cite{DBLP:journals/corr/abs-1808-03156}. Apple devices always send out \gls{BLE} advertisements before they setup a peer-to-peer connection over \gls{AWDL}, e.\,g., for exchanging files.
In the future we want to visualize those connections over \gls{AWDL} by monitoring both interfaces in parallel. 

If Bluetooth chips adopt \gls{AoA} support and it becomes available through public APIs, \appname{} could leverage the \gls{AoA} of wireless signals to align devices in the proximity view in \cref{sec:env_scanner} with actual angles.
Finally, as our core framework is written in Swift, portability to other platforms such as Linux is possible with some caveats.

\begin{figure}
    \centering
    \includegraphics[width=\linewidth]{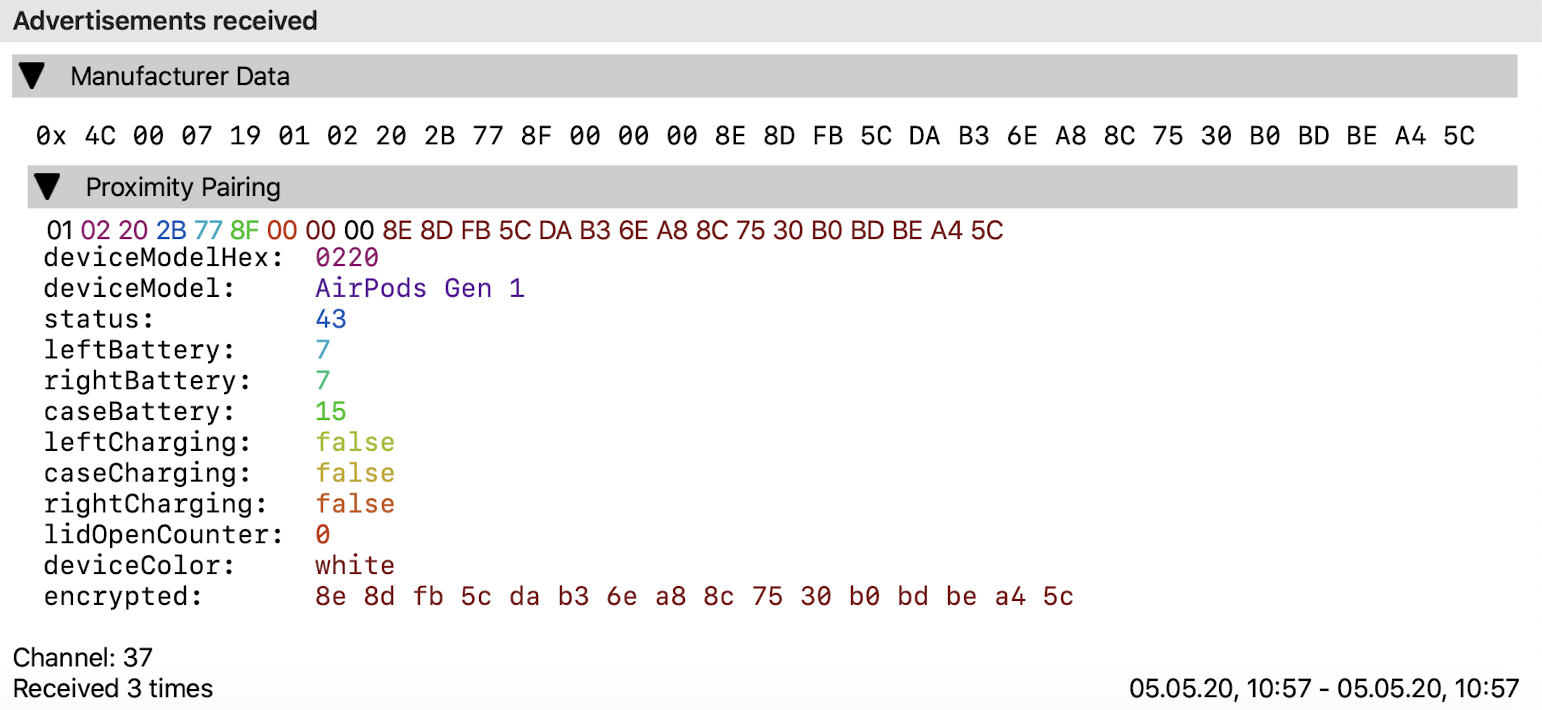}
    \caption{Decoding of an BLE advertisement emitted by Apple \emph{AirPods} earphones}
    \label{fig:airpods_ble}
\end{figure}

\section{Demonstration setup}


Our demonstration provides every participant with an application binary running on macOS and gives them the ability to join a TestFlight Beta for an iOS build. We provide an introduction to \appname{} and present the available features during a live session or via a pre-recorded video. 
The demonstration starts with the general scanning and dissection features: we use Apple AirPods advertisements to detect the device model and the state, as shown in \cref{fig:airpods_ble}. We also demonstrate Apple AirDrop advertisements.
We continue the demonstration by showing the distance estimations when devices are moving around the scanning device. Then we use the scanning device to find hidden \gls{BLE} devices. 
Finally, we show how to enumerate devices in combination with a Raspberry Pi as an external scanning source.
Participants can test the software immediately and report any recommendations for future releases on GitHub.


\section*{Availability}

\appname{} is available open-source on \url{https://github.com/seemoo-lab/BTLEmap}. All necessary frameworks are also published on our GitHub account. 
Furthermore, the app will be made available on TestFlight as soon as possible.
After our Beta test, we will distribute \appname{} via the official Apple App Store for free. It allows other researchers without a development background to get easy access to our tool and it includes non-scientific users that are curious about their \gls{BLE} environment.

\section*{Acknowledgements}
This work has been funded by the German Federal Ministry of Education and Research and the Hessen State Ministry for Higher Education, Research and the Arts within their joint support of the National Research Center for Applied Cybersecurity ATHENE.

\bibliographystyle{ACM-Reference-Format}
\bibliography{main.bib}

\end{document}